# Observation of asymmetric solitons in waveguide arrays with refractive index gradient

Steffen Weimann,[1] Yaroslav V. Kartashov,[2,3,*] Victor A. Vysloukh,[2] Alexander Szameit,[1] and Lluis Torner[2]

[1]Institut für Angewandte Physik, Friedrich-Schiller-Universität Jena, Max-Wien-Platz 1, 07743 Jena, Germany
[2]ICFO-Institut de Ciencies Fotoniques, and Universitat Politecnica de Catalunya, Mediterranean Technology Park, 08860 Castelldefels (Barcelona), Spain
[3]Institute of Spectroscopy, Russian Academy of Sciences, Troitsk, Moscow Region, 142190, Russia

We study light propagation in waveguide arrays made in Kerr nonlinear media with a transverse refractive index gradient, and we find that the presence of the refractive index gradient leads to the appearance of a number of new soliton families. The effective coupling between the solitons and the localized linear eigenmodes of the lattice induces a drastic asymmetry in the soliton shapes and the appearance of long tails at the soliton wings. Such unusual solitons are found to be completely stable under propagation, and we report their experimental observation in fs-laser written waveguide arrays with focusing Kerr nonlinearity.

Formation of self-sustained nonlinear waves in optical media with a transversely periodic shallow refractive index modulation (or optical lattices [1]) has been explored and observed in various physical settings. Spatial lattice solitons have been studied in semiconductor waveguide arrays [2], in optically induced [3,4] or technologically fabricated [5] arrays in photorefractive media, in voltage-controlled arrays in liquid crystals [6], fiber arrays [7] and in fs-laser written waveguide arrays in fused silica [8], to name just a few examples. Thus, today the properties of solitons in standard lattices are well understood (see the reviews [9,10] and references therein). A common feature of such solitons is that they may be viewed as nonlinear defect modes, whereby their propagation constants fall into forbidden gaps of the Floquet-Bloch spectrum of the array, while in the low-power limit all states transform into periodic Bloch waves, namely eigenmodes of the perfectly periodic lattice.

However, the picture changes drastically if on top of the shallow periodic modulation the refractive index grows linearly in the transverse plane. In analogy with the behavior of electronic wavefunctions in combined periodic and dc fields [11], the linear eigenmodes of such waveguide arrays become localized and their propagation constants form a so-called Wannier-Stark (WS) ladder [12]. Due to the corresponding equidistant set of allowed propagation constants, linear excitations periodically restore their shape after a certain propagation distance, even after a considerable initial reshaping. Such Bloch oscillations have been observed in both one- and two-dimensional waveguide arrays [13-18].

On the one hand, the presence of a weak nonlinearity is known to destroy the Bloch oscillations [15,19,20], and on the other hand, nonlinearity may lead to the formation of solitons. In waveguide arrays with a refractive index gradient the properties of the solitons differ considerably from the soliton properties in homogeneous arrays, due to a principally different behavior of the excitations in the low-power limit. So far, solitons in non-homogeneous discrete lattices have been studied only in discrete media with focusing Kerr nonlinearity [21,22] and, to the best of our knowledge, they have not been observed experimentally.

In this Letter we address the properties of lattice solitons in arrays with a transverse refractive index gradient in the presence of focusing or defocusing nonlinearity. We find that, in contrast to homogeneous arrays, there are several distinct, new families of solitons. After summarizing the nonlinear eigenmodes of the system, we study the light evolution dynamics in the inhomogeneous arrays. The Bloch oscillations that exist in the arrays for low input powers are destroyed and rich dynamical phenomena appear as the power is increased. We report the experimental observation of such phenomena, including the destruction of the Bloch oscillations, in a femtosecond laser-written waveguide array. First we will present our results in media with focusing nonlinearity while the last part of the Letter is devoted to the results obtained in media with defocusing nonlinearity.

We describe the propagation of a light beam in a Kerr-type medium with a transversally, linearly growing refractive index by the nonlinear Schrödinger equation

$$i\frac{\partial q}{\partial \xi} = -\frac{1}{2}\frac{\partial^2 q}{\partial \eta^2} + \sigma q |q|^2 - pR(\eta)q - \alpha\eta q, \qquad (1)$$

Here $\xi$ is the propagation distance normalized to the diffraction length $kx_0^2$ of the dimensionless electric field $q = (k^2 x_0^2 n_2/2n)A$; $\eta$ is the transverse coordinate normalized to the characteristic transverse scale $x_0 = 10$ $\mu$m. The constant $p$ stands for the waveguide parameter $p = \delta n k^2 x_0^2/n$, where $\delta n$ is the amplitude of the refractive index modulation. The focusing or defocusing nature of the nonlinearity corresponds to $\sigma = -1$, $\sigma = +1$, respectively. The slope of the transverse refractive index gradient is dictated by the parameter $\alpha$. The function $R = \sum_k G(\eta - kd)$ describes the underlying homogeneous refractive index pro-file of the array, consisting of a set of super-Gaussian waveguides $G(\eta)=\exp(-\eta^6/w_\eta^6)$ of width $w_\eta$ separated by the distance $d$. In accordance with the actual experimental parameters, we set $a=0.3$ (this corresponds to 3 $\mu$m-wide waveguides), $d=3.2$ (32 $\mu$m period of the array), and $p=2.6$ (refractive index modulation depth $\delta n \approx 2.9\times 10^{-4}$). At the wavelength $\lambda = 800$ nm, a sample length of 100 mm in a material with refractive index $n=1.46$ used in the experiments corresponds to $\xi = 87.2$. The transverse refractive index gradient $\alpha = 0.0113$ was selected in such a way that the longitudinal period of Bloch oscillations $\xi_B = 2\pi/\alpha d$ corresponds to 200 mm. The amplitude of Bloch oscillations can be estimated as $\eta_B \approx \delta b/\alpha$, where $\delta b$ is the width of the first allowed band in the Floquet-Bloch spectrum of the periodic structure [23].

First, we consider a focusing medium. We look for stationary solutions of Eq. (1) in the form $q(\eta,\xi) = w(\eta)\exp(ib\xi)$, where $b$ is the propagation constant. Linear stability analysis is performed by substituting perturbed solutions in the form $q(\eta,\xi) = [w(\eta) + u\exp(\delta\xi) + iv\exp(\delta\xi)]\exp(ib\xi)$ with $u,v \ll w$, into Eq. (1). After linearization of the remaining equation for the perturbations $u,v$ one obtains the following eigenvalue problem

$$\delta u = -\frac{1}{2}\frac{\partial^2 v}{\partial \eta^2} + bv + \sigma vw^2 - pRv - \alpha\eta v,$$
$$\delta v = +\frac{1}{2}\frac{\partial^2 u}{\partial \eta^2} - bu - 3\sigma uw^2 + pRu + \alpha\eta u.$$
(2)

The dependence of the dimensionless energy flow $U = \int_{-\infty}^{\infty} w^2 d\eta$ on $b$ for all obtained soliton families is shown in Fig. 1(a), and illustrative examples of soliton shapes are depicted in Fig. 1(c)-(d). Whereas in a perfectly periodic array one obtains only one family of odd solitons bifurcating from the top of the first allowed band [dashed line in Fig. 1(a)], *in the array with the linear refractive index gradient one encounters multiple new soliton families*, including lower and upper branches joining at a cutoff value of $b$. Only for the lowest soliton branch representing a nonlinear modification of the WS mode centered at $\eta=0$ (leftmost curve) the slope $dU/db$ is positive for all $b$ values. In Fig. 1(a), we indicate two specific solitons of this zero branch using circles. Their spatial profiles are shown in Fig. 1(c). Since the soliton family bifurcates from the WS mode, the soliton remains localized when $b$ approaches its cutoff value. It should be stressed that such cutoff is smaller than the cutoff of the soliton family in the gradientless array (i.e., the bandgap spectrum at $\alpha=0$ cannot be used for the prediction of the soliton existence domain).

The solitons feature strongly asymmetric shapes – they exhibit multiple oscillations at the right wing (the width of the solution in the low-power limit is determined by the refractive index gradient $\alpha$). When $U$ increases, the soliton amplitude increases and its width grows as the oscillating wings at $\eta>0$ become more significant. For large values of $U$, the wings acquire a nearly parabolic envelope. Solitons from all other branches are composite states involving a main spike at $\eta=0$ effectively coupled to the nonlinear WS modes centered on various lattice sites located at $\eta = kd$, $k>0$.

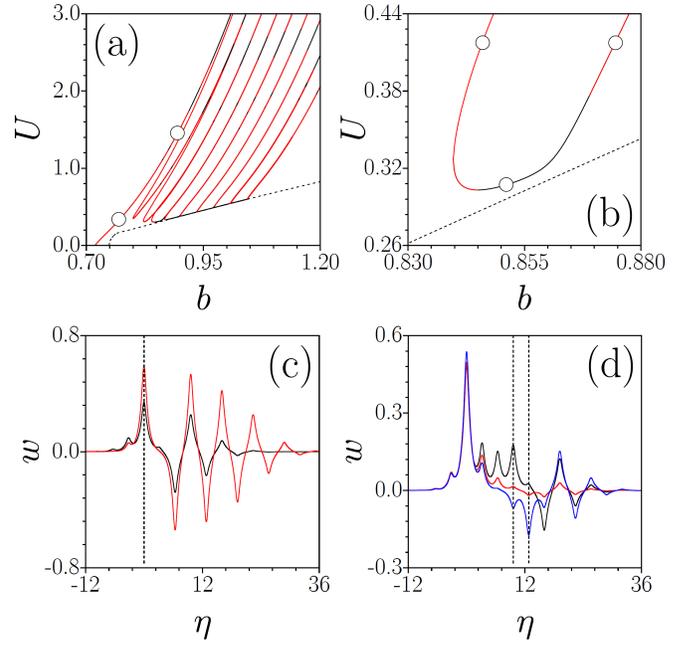

Fig. 1. (a),(b) Energy flow versus propagation constant for solitons in a focusing medium. Panel (b) shows the magnified third branch. Stable branches are shown in black, while unstable ones are shown in red. The dashed lines stand for the soliton existing in waveguide array without refractive index gradient. Solitons in focusing media (c) from zero branch at $b=0.770$ (black), $b=0.906$ (red) and (d) from third branch at $b=0.846$ (black), $b=0.851$ (red), and $b=0.875$ (blue). The soliton profiles depicted in (c),(d) correspond to the circles in (a) and (b). Vertical dashed lines indicate the centers of the WS modes resonantly coupled with the main peak at $\eta=0$ or giving rise to an entire soliton family.

The location of the WS mode that effectively couples with the main spike determines the branch number and the propagation constant cutoff at which the branch emerges. Cutoffs are directly connected to the equidistant eigenvalues of the linear WS modes, as readily visible in Fig. 1(a), where new branches emerge with nearly equal steps in $b$. Figure 1(d) shows the profiles of the solitons corresponding to the circles in Fig. 1(b). Solitons from the upper part of this third branch (black line) are an in-phase combination of the central spike with the WS mode centered on the third waveguide. The soliton from the lower part of this branch with the same $U$ value (blue line) involves the out-of-phase central spike and the WS mode from the fourth waveguide. The soliton from the lower branch close to the cutoff (red line) is well localized. It represents an intermediate state with a well pronounced central spike and a low-amplitude tail. As one can see from Fig. 1(a), the energy flow of such soliton is close to that of the soliton in the array with $\alpha=0$.

As the branch number increases, the soliton tail arising due to the coupling with the WS mode, becomes less and less pronounced, at least close to the $b$ value where the branch emerges. In the experiments, described below, we will excite one of these well-localized states. However, for a given branch the right tail abruptly grows and acquires a characteristic parabolic envelope if one follows the lower or upper parts of the branch into the region of high $U$ values.

Figures 1(a),(b) also indicate the stability of the soliton solutions according to their perturbation growth rate $\delta$. The real part of this perturbation growth rate is a measure of how fast the intensity distribution diverges from the stationary soliton solution after a small perturbation of the field. If the real part of the growth rate is zero, the soliton is stable. Regions of stable solitons are shown in black, while regions that correspond to unstable solutions are shown in red in the plots. Not only well-localized solitons from the bottom of the lower branch can be stable, but also multiple stability domains appear at high $U$ values for strongly asymmetric solitons with long right oscillating wings. For sufficiently high values of $U$, all solitons from the lower branch become stable [Fig. 1(a)]. In contrast, solitons from the upper branches are always exponentially unstable.

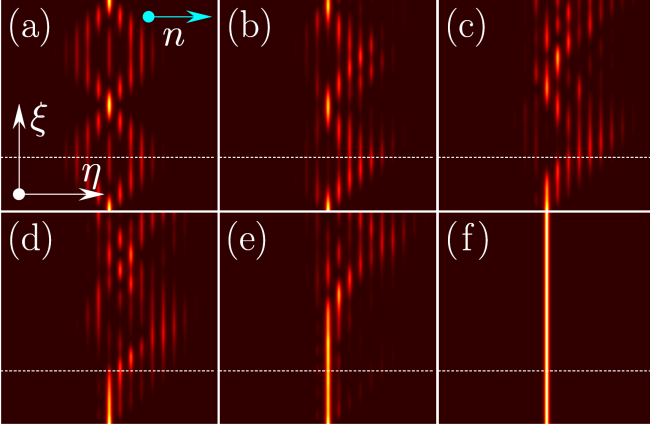

Fig. 2. Destruction of the Bloch oscillations and excitation of solitons in an array with a refractive index gradient in focusing nonlinear media. The input continuous wave amplitude is (a) $A = 0.052$, (b) $0.424$, (c) $0.545$, (d) $0.568$, (e) $0.586$, and (f) $0.944$. Horizontal dashed lines indicate the length of the sample used in our experiments. The blue arrow shows the direction of refractive index growth. (f) corresponds to the excitation of the soliton with $U \approx 0.88$, $b \approx 1.23$ from one of the stable lower branches.

We also studied the excitation of the above mentioned solitons with single-channel inputs $q(\eta, \xi = 0) = A w_{\text{lin}}(\eta)$, where $w_{\text{lin}}(\eta)$ describes the profile of the eigenmodes of the isolated waveguide, while $A$ is its amplitude. Figure 2 illustrates the gradual transition from Bloch oscillations at small values of $A$ [Fig. 2(a)] to the formation of a well-localized soliton at large input amplitudes in the nonlinear focusing medium [Fig. 2(f)]. Increasing the nonlinear contribution, results in a notable asymmetry of the Bloch oscillations. Up to $A = 0.310$, oscillations remain nearly periodic, but they shift in the direction of increasing refractive index [Fig. 2(b)]. Any further increase of the input amplitude makes the oscillations irregular [Figs. 2(c),(d)], and at one point (already at $A > 0.6$) one observes an abrupt localization [Fig. 2(f)].

To experimentally observe the predicted phenomena, we inscribed waveguide arrays into fused silica via femtosecond laser writing [24]. This allows us to study the nonlinear evolution dynamics in arrays with focusing nonlinearity. Observation of the light distribution is accomplished by imaging the light intensity at the output facet of the sample with a CCD camera. The resulting picture contains a cross section of the intensity distribution at $\xi = \xi_{\text{out}}$. In order to observe the Bloch oscillations at low input powers, one needs a linear increase of the propagation constants from one end of the array to the other. Although the propagation constants of the waveguides could be engineered by changing the laser writing parameters, a quadratic bending of the whole array in the $(\eta, \xi)$-plane is applied to provide the intended effect on the light propagation [25]. In other words, we adjusted the slope $a$ of the refractive index by the curvature of the array as a whole and not by adjusting the laser parameters for every single waveguide. Approximating the quadratic trajectory $\xi = a\eta^2$ of each waveguide by a circle, the radius of curvature $r$ and the length of one Bloch period $\xi_B$ are connected via $r = \xi_B n d / \lambda$. On the one hand, our samples are $10$ cm-long and on the other hand, we want to observe the effect of nonlinearity right at one half of a Bloch period. Hence the radius of curvature was set to $r = 11.68$ m in the fabricated array.

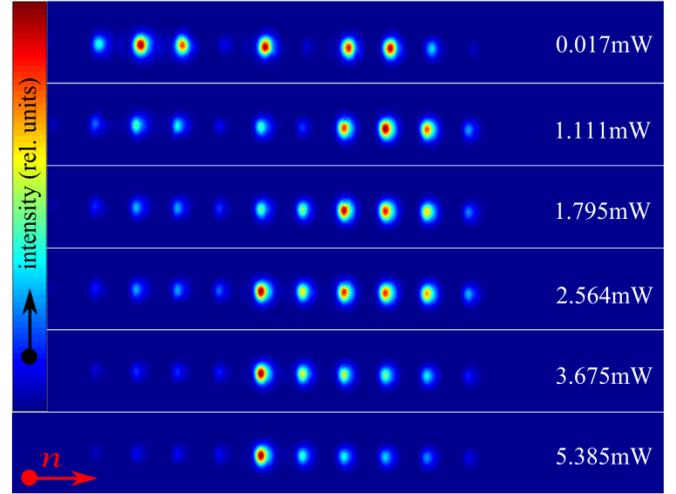

Fig. 3. Experimental output intensity distributions at different peak power levels illustrating the transition from Bloch oscillations to soliton formation. The corresponding calculated evolution dynamics is shown in Fig. 2. The red arrow indicates the direction of refractive index growth.

In Fig. 3 we present the observed intensity distributions at the output facet of the fabricated array, for six input powers ranging from $0.01$ mW to $5.39$ mW. The power could not be increased further without exceeding the damage threshold of the material. The asymmetric shape of the intensity distribution as well as the formation of a localized spot at high input powers are clearly visible. For an input power of $5.4$ mW, the central waveguide contains only 32% of the total power, while in simulations the power concentrates in the central guide. However, our theoretical model deals with continuous waves, while in experiment we use 300fs pulses. The wings of pulses always provide diffracting background. At all input powers, a good agreement between the predicted evolution dynamics and the measured impact of the nonlinearity is obtained.

We also studied theoretically self-defocusing media and we found a similarly rich structure of soliton families [see Figs. 4(a),(b)]. Because self-defocusing nonlinearity leads to a decrease of the propagation constant, now the central spike couples to WS modes having lower propagation constants and residing at $\eta < 0$ [a representative example of such a coupling is shown in Fig. 4(c)]. Notice that increasing the energy flow is accompanied by the development of a long left tail with in-phase field oscillations [Fig. 4(c)]. This is in contrast to the case of focusing nonlinearity described above, where oscillations on the growing right soliton wing are out-of-phase [Fig. 1(c)]. Again, the set of soliton branches in-cludes one branch without any threshold bifurcating from the WS mode centered at $\eta=0$ and an infinite number of branches with energy flow thresholds and nearly equidistant, progressively decreasing cutoffs.

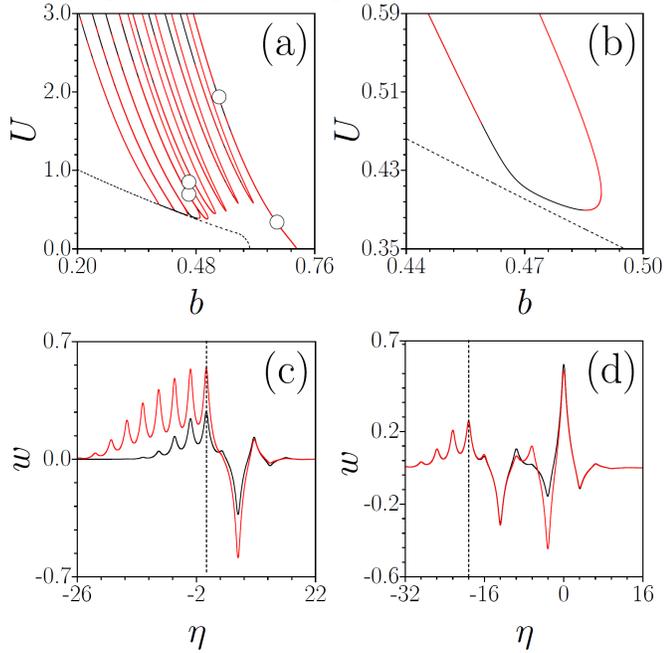

Fig. 4. (a) Energy flow versus propagation constant for solitons in a defocusing medium. (b) Enlarged portion of $U(b)$ dependence for the sixth branch. The meaning of black, red, and dashed lines is the same as in Fig. 1. Solitons (c) from zero branch at $b=0.670$ (black), $b=0.534$ (red), and (d) from the lower and upper parts of the fifth branch at $b=0.463$ corresponding to the circles in (a).

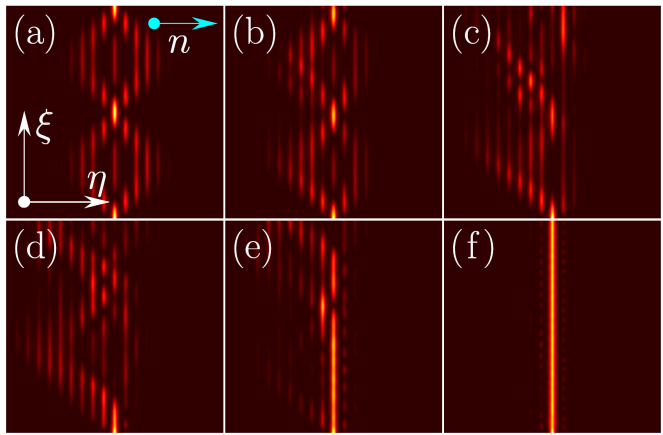

Fig. 5. Predicted destruction of Bloch oscillations and excitation of solitons in an array with refractive index gradient in a self-defocusing nonlinear medium. The input amplitude is (a) $A=0.052$, (b) $0.424$, (c) $0.545$, (d) $0.696$, (e) $0.721$ and (f) $0.944$.

In Figs. 4(a),(b) we show regions of stable and unstable soliton solutions, in black and red, respectively. We see that from the 0-th to the 5-th branch in defocusing media all solitons with a small energy flow are unstable. Moreover, the regions of stability occur very irregularly in Fig. 4(a). Since the majority of soliton solutions extents into the negative $\eta$-direction [Fig. 4(d)], the evolution dynamics after a single waveguide excitation changes accordingly. The dynamical behavior is summarized in Fig. 5. The Bloch oscillations ex- perience a displacement in the direction opposite to the refractive index gradient. As in a focusing medium, the effect of nonlinearity is to destroy the Bloch oscillations and soliton formation occurs above a certain energy flow threshold.

Summarizing, we studied the shape and the dynamic stability of new soliton families existing in waveguide lattices with a transverse refractive index gradient in nonlinear media. We observed the predicted phenomena using femtosecond laser-written waveguide arrays. Observations confirmed the soliton asymmetry and destruction of the Bloch oscillations. The complex shapes of the solitons arise from the Wannier-Stark modes of the linear regime.